# IBIC analysis of CdTe/CdS solar cells.

E. Colombo[1,2], A. Bosio[3], S. Calusi[1,4], L. Giuntini[4], A. Lo Giudice[1,2], C. Manfredotti[1,2], M. Massi[4], P. Olivero[1,2], A. Romeo[5], N. Romeo[3] and E. Vittone[1,2]■

[1] INFN, sezione di Torino, via P. Giuria 1, 10125, Torino, Italy.

[2] Experimental Physics Dept. and NIS excellence centre, University of Torino, via P. Giuria 1,10125, Torino, Italy.

[3] Physics Department, University of Parma, v.le G.P. Usberti 7/A, 43100, Parma, Italy.

[4] Physics Dept. University of Firenze and INFN Sezione di Firenze, Via Sansone 1, 50019, Sesto Fiorentino, Firenze, Italy.

[5] Science & Technology Dept. University of Verona, Ca Vignal 2, Strada Grazie 15, I-37134, Verona, Italy.

## Abstract

This paper reports on the investigation of the electronic properties of a thin film CdS/CdTe solar cell with the Ion Beam Induced Charge (IBIC) technique. The device under test is a thin film (total thickness around 10 $\mu$m) multilayer heterojunction solar cell, displaying an efficiency of 14% under AM1.5 illumination conditions. The IBIC measurements were carried out using focused 3.150 MeV He ions raster scanned onto the surface of the back electrode. The charge collection efficiency (CCE) maps show inhomogeneous response of the cell to be attributed to the polycrystalline nature of the CdTe bulk material.
Finally, the evolution of the IBIC signal vs. the ion fluence was studied in order to evaluate the radiation hardness of the CdS/CdTe solar cells in view of their use in solar modules for space applications.

---

■ corresponding author: Ettore Vittone, e-mail: vittone@to.infn.it

**Introduction**

Thin film solar cells have the great advantage of being much easier to produce compared to crystalline silicon solar cells, giving a very high throughput with halved crystallization temperatures and monolithical interconnection.

In particular CdTe is particularly suitable for industrial mass production since it grows stoichiometrically with substrate temperatures above 250 °C and can be successfully deposited with a large variety of techniques such as vacuum evaporation (VE), close space sublimation (CSS), vapor transport deposition (VTD), RF-sputtering, electro-deposition, screen printing, etc [1].

CdTe bandgap of 1.45 eV is very near to the theoretical maximum conversion efficiency (31%), while open circuit voltage and short circuit current are maximized. These substantial advantages have been proven by the impressive increase of CdTe photovoltaic modules production from a few MW in 2003 up to over 1 GW in 2008 with constantly decreasing production costs. Moreover, it has been shown that CdTe has the highest stability under proton and electron irradiation compared to the other photovoltaic devices, which makes CdTe cells very promising for space applications [2] .

The CdTe solar cells can be divided in four different sections [1]. The front contact, which is the first layer deposited directly on the glass substrate, consisting of a transparent conducting oxide (TCO), typically indium tin oxide (ITO) or tin oxide doped with fluorine (FTO); the CdS (also known as window layer), which is the n-type semiconductor of the junction and has to be optically transparent in order to allow the absorber to convert all the light spectrum; the CdTe, which has the double function of absorbing the light (giving place to the electron-hole pair generation) and to be the p-type semiconductor of the junction. Due to the high absorption coefficient of CdTe for photons in the visible range, the thickness of this layer is usually less than 10 $\mu$m. Finally the back contact, typically a metal/metal or semiconductor/metal bi-layer.

In this paper we present the first IBIC characterisation of such thin film solar cell. The first motivation of this research is to demonstrate the feasibility of the IBIC technique to investigate the electronic properties of such thin film devices and to map their charge collection efficiency. The material is in fact polycrystalline and carrier traps and the electric field associated with grain boundaries will affect the diffusion of the carrier excess in the CdTe layer. With respect to other studies carried out with electron beam induced current measurements (EBIC) [3], the novelty of the use of energetic ions consists in the analysis of the bulk properties of the finite device.

Secondly, due to the interesting perspective of using such cells for space applications, we performed an investigation on the radiation hardness of cells under 3.150 MeV He ion bombardment through the continuous monitoring of the CCE vs. the ion fluences.

**Experimental**

The scheme of the multilayer CdS/CdTe cell that was fabricated by the Parma University group is shown in Fig. 1. The transparent contact is 500 nm indium tin oxide deposited by DC-reactive sputtering at a substrate temperature of 300 °C on soda lime glass; a 150 nm layer of zinc oxide is deposited by RF-sputtering at a substrate temperature of 200 °C. The n-type material of the heterojunction is constituted by a 150 nm CdS layer which is deposited by RF-sputtering; the p-type side consists of a 8 $\mu$m thick CdTe layer deposited by close space sublimation at a temperature of 500 °C. The junction is then activated by a recrystallization treatment of the CdTe, typically a "CdCl$_2$ treatment" that in our case consists in an annealing at 400 °C in freon atmosphere. Finally the metal contact is a bi-layer of As$_2$Te$_3$:Cu (which gives the ohmic contact with CdTe) and Mo layer. The finished devices have typical conversion efficiencies between 14 and 15 % [4]. A SEM image of the cell is shown in Fig. 2.

The lateral dimensions of the cell under investigation are (4.4x4.4 mm$^2$).

The current voltage characteristic is shown in Fig. 3. At low bias voltage, the reverse current is of the order of fractions of $\mu$A, which induces an high noise level at the input of the preamplifier. For this reason, IBIC measurements were carried out without any applied bias voltage.

Silver paste was used to bond an Au wire for the connection of the Mo electrode to an Amptek A250 charge sensitive preamplifier and the ITO transparent electrode to ground. The charge signal was amplified with a shaping amplifier (from Silena 761) and fed into an OM-DAQ acquisition system. At zero bias the capacitance of the cell is more than 100 pF (see the C-V curve in Fig. 4) which corresponds to an equivalent noise energy of about 88 keV as evaluated from the FWHM of the signal from a precision pulse generator.

The IBIC measurements were performed at the external scanning ion microprobe facility installed at the 3 MV Tandem Accelerator [5] at LABEC in Florence (I).

A 3150 keV He beam was extracted in air through a silicon nitride ($Si_3N_4$) window (50 nm thick, 1 mm$^2$ area) and focused and raster scanned by means of Oxford Microbeam system over the Mo surface of the cell. The window is inclined of 60° with respect to the beam axis, and the sample was located as close as possible to the exit window (Fig.5). The measurement were performed in air; the energy of the ions transmitted through the $Si_3N_4$ window and the 1 mm thick air layer interposed between the sample and the centre of the exit window was 3000 keV with a FWHM of 11 keV and the lateral straggling of the beam was about 4 $\mu$m in FWHM, as evaluated by SRIM simulation.

**Results and discussion**

Fig. 6 shows an IBIC map of the CdTe/CdS solar cell. The black region on the top side is due to the silver paste used for bonding. In average, about 20 pulses were collected at each pixel and the map was obtained encoding the average charge pulse height in a grey scale; the histogram of the pixel distribution vs. the pulse height is shown in Fig. 7, as resulting from the elaboration of the total scan recorded in event-by-event mode. It is worth noticing that such histogram is much more peaked than the total pulse spectrum shown in the same figure. Two facts can be used to explain the difference of the two distributions. The first concerns the smoothing of the pulse dispersion due to the energy straggling of the incident ions. Being such dispersion symmetric, the average method allows the determination of a single representative pulse height, which is relevant to the mean energy of the ions. The second fact is relevant to the statistics of carriers contributing to the induced charge. Being the device partially depleted, a large fraction of the pulse height derives from carriers generated outside the depletion region as demonstrated in Fig. 8, where the generation profile from He ions of energy 3 MeV on the multilayer cell is reported. These carriers contribute to the induced charge because of their injection by diffusion from the neutral region to the active region (i.e. where the electric field occurs) [6]. Being the diffusion process much slower of the drift transport, these carriers are more prone to recombination and trapping, resulting in a statistical fluctuation of the number of charge arriving at the device junction, where they induce the IBIC signal together with the drift charge. Hence, the average method allows the determination of a single value which is representative of the average number of carriers arriving at the depletion region.

The CCE map reported in Fig. 6 clearly shows non-homogeneities to be attributed to the polycrystalline nature of the CdTe layer. In the zoomed (100x100 $\mu m^2$) region reported in the inset, structures showing higher CCEs have dimensions of the order of 10 $\mu m$, surrounded by darker (i.e. low efficiency) boundaries.

Since the dimension of the features in the IBIC maps is comparable to the size of the CdTe microstructures after $CdCl_2$ re-crystallization, the attribution of the dominant role of grain boundaries in the limitation of the carrier drift lengths appears very convincing. In fact, in analogy with what has been extensively reported after IBIC studies on polycrystalline CVD diamond [7] and, recently, on polycrystalline CdTe radiation detectors [8], grain boundaries act as preferential regions where recombination or trapping occurs, and the contrast in the IBIC image highlights grain shapes in the bulk material.

In order to evaluate the sensitivity of the cell to ion damage, a 100x100 $\mu m^2$ area was scanned by the He beam up to a total fluence of $2 \times 10^{10}$ ions·cm$^{-2}$. After this selective irradiation, this area presents a lower charge collection efficiency with respect to the pristine regions, as shown in Fig. 9. Clearly, the ion bombardment induces damage in the material, generating recombination centres, resulting in a decrease of the induced charge signal. In order to analyse the variation of the charge pulse height with cumulative fluence, IBIC spectra and maps were taken at regular fluence intervals during the irradiation. From the analysis of the spectra a decrease of the average pulse height vs. the ion fluence is evident. This effect is represented in the spectra (obtained by the elaboration of 2000 pulses uniformly distributed onto the irradiated area) of Fig. 10 at four different fluences, and summarised in Fig. 11, where the average pulse height is reported as function of ion fluence. In the same graph the behaviour of the standard deviation of the pulse height distribution is also reported, which highlights the increasing uniformity (i.e. decreasing dispersion) of the response as the fluence increases. This is further confirmed by the maps in Fig. 12. High efficiency grains observable after small ion fluences disappear in the maps relevant to more damaged regions, which present lower contrast.

**Conclusions**

This paper presents the first IBIC characterisation of a CdTe/CdS multilayer solar cell. The experiments were carried out using a 3.150 He ion beam extracted in air and focused onto the back electrode of the solar cell. Despite the employment of an extracted beam in air, the spatial resolution of the IBIC maps was proven to be adequate to resolve the polycrystalline structure of the bulk materials, highlighting the role played by grain boundaries to limit the drift lengths of carriers.

Damage effects induced by ion irradiation was evaluated by analysing the average IBIC signal coming from a 100x100 $\mu m^2$ region irradiated up to a fluence of $2 \cdot 10^{10}$ ions/cm$^2$. The average pulse height decreases of 20% with respect to the pristine case; the analysis of the maps recorded at different fluences shows that the decreasing of CCE is not uniform; in fact large grains, showing higher efficiencies in the pristine case, are more sensitive to radiation effects. As a consequence, ion irradiation levels towards low CCE values the response of the cell, reducing the spatial fluctuations as indicated microscopically by the decreasing contrast of IBIC maps.

**Acknowledgements**

This work has been carried out and financially supported in the framework of the INFN experiment "DANTE".

**Figure Captions**

Fig. 1   Scheme of the CdTe/CdS cell, with the relevant electrical connections.

Fig. 2   SEM image of the surface of the solar cell.

Fig. 3   Current-voltage characteristic of the CdTe/CdS solar cell.

Fig. 4   C-V and $1/C^2$-V curves of the CdTe/CdS solar cell.

Fig. 5   Scheme of the exit window – sample geometry. The dimension of the $Si_3N_4$ window is 1x1 mm$^2$.

Fig. 6   IBIC map of the CdTe/CdS solar cell. In the inset, a zoom of the map (100x100 $\mu m^2$)

Fig. 7   Pulse height frequency histogram (right vertical scale) and pixel distribution (left vertical scale) relevant to the map in the previous figure. The pixel distribution was evaluated considering a pulse threshold at channel 150.

Fig. 8   Simulated energy loss profile in the multilayer cell generated by 3 MeV He ions.

Fig. 9   CCE map of the solar cell after the selective irradiation of the area highlighted in the inset in Fig. 6.

Fig. 10  Spectral evolution of the average IBIC signals recorded after different ion fluences. The measured charge was evaluated by normalising the pulse response by a Si barrier detector used as reference.

Fig. 11  average CCE (left scale, full square) and standard deviation (right scale, open circle) vs. ion fluence. Data are normalised to the pristine case.

Fig. 12  IBIC maps of the irradiated regions after different fluences.

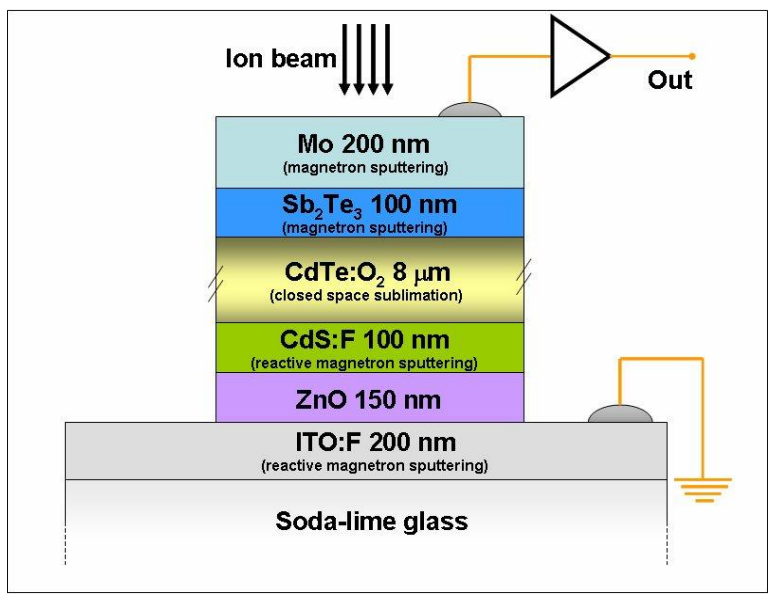

Fig. 1  Scheme of the CdTe/CdS cell, with the relevant electrical connections.

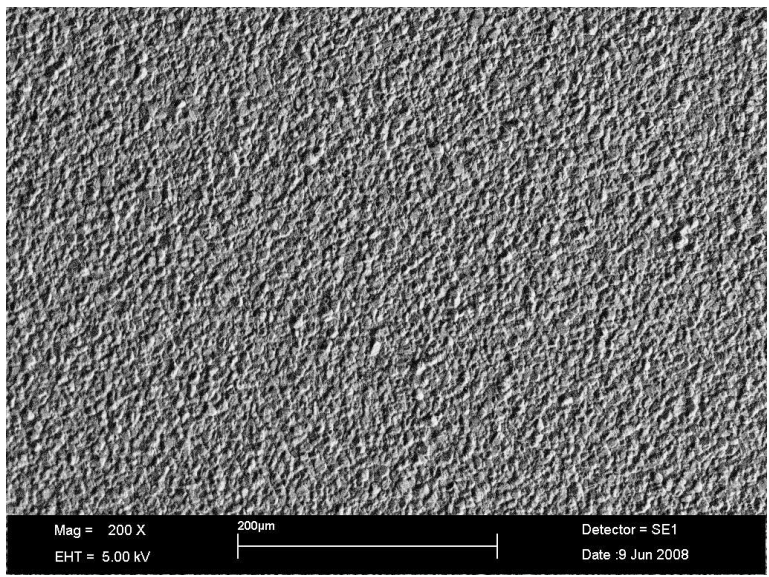

Fig. 2  SEM image of the surface of the solar cell.

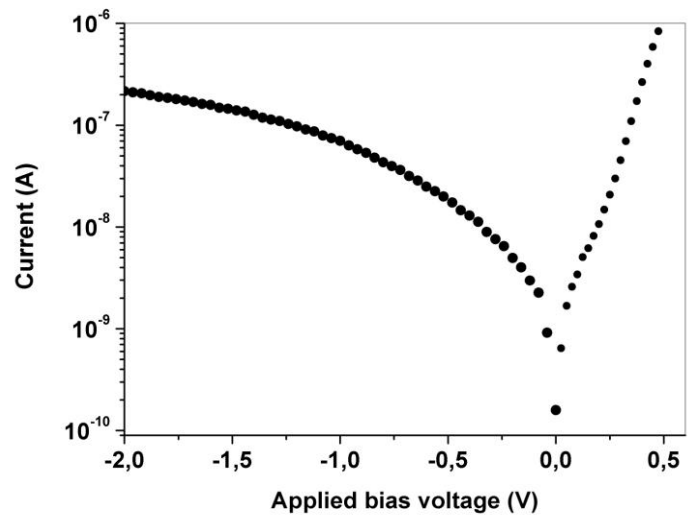

Fig. 3   Current-voltage characteristic of the CdTe/CdS solar cell.

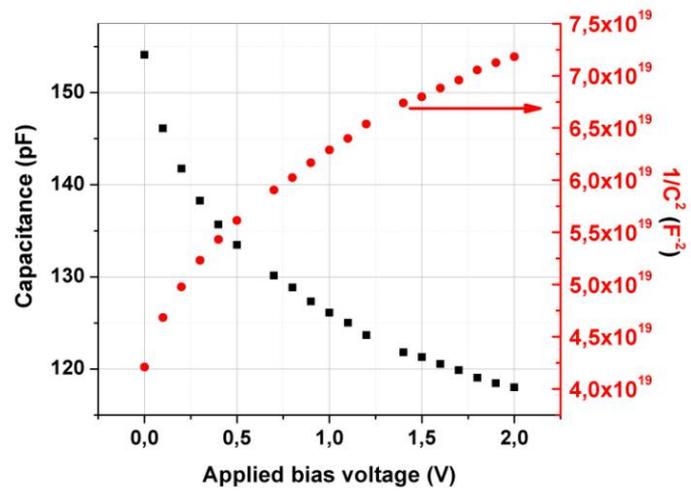

Fig. 4  C-V and 1/C$^2$-V curves of the CdTe/CdS solar cell.

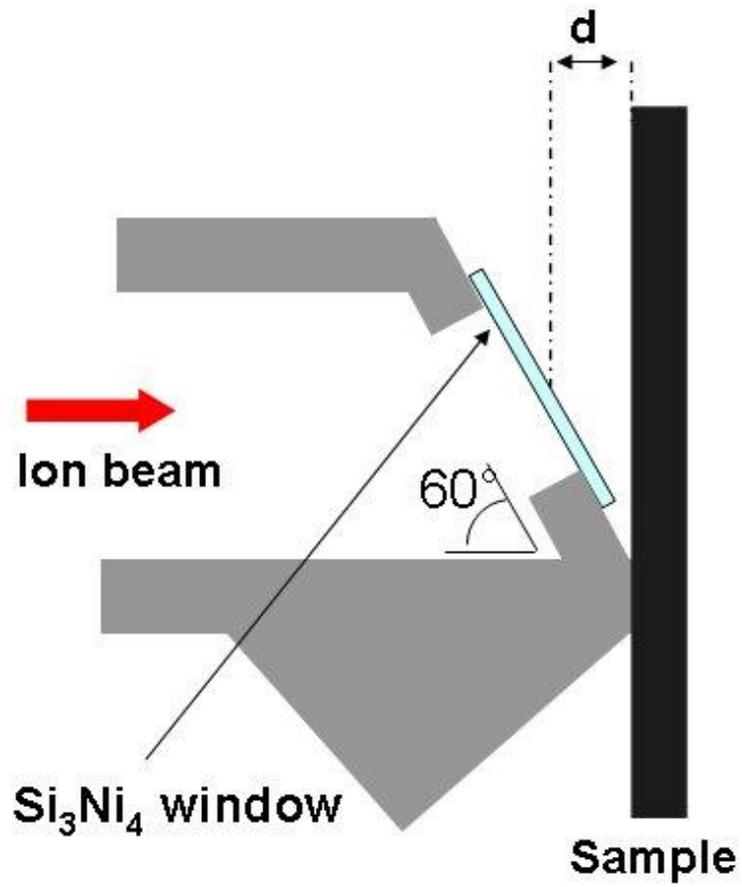

Fig. 5 Scheme of the exit window – sample geometry. The dimension of the Si$_3$N$_4$ window is 1x1 mm$^2$.

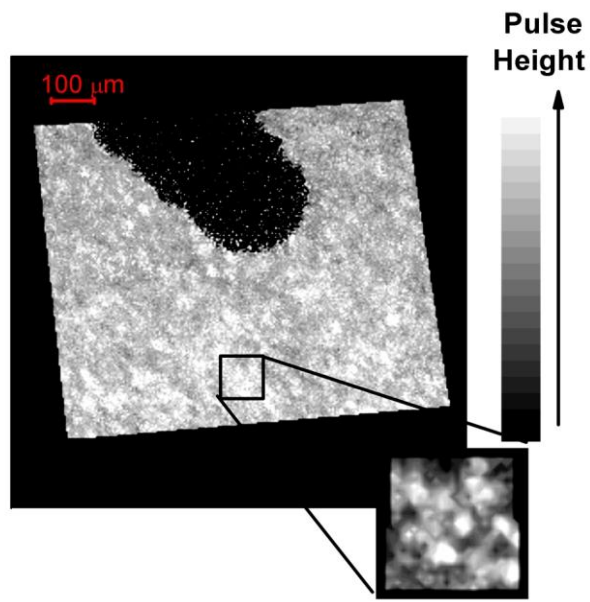

Fig. 6  IBIC map of the CdTe/CdS solar cell. In the inset, a zoom of the map (100x100 $\mu m^2$)

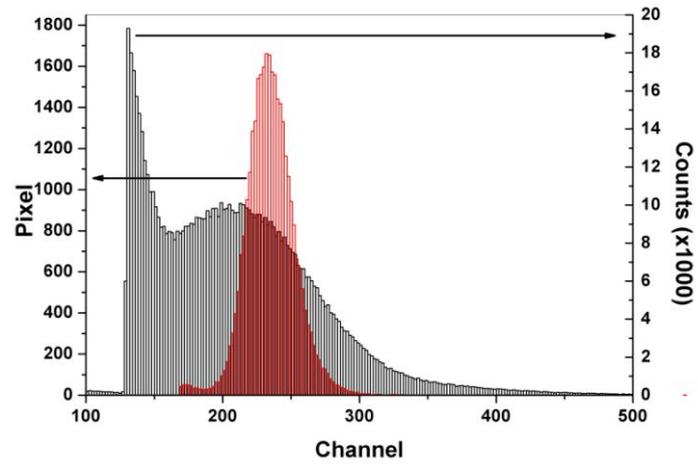

Fig. 7 Pulse height frequency histogram (right vertical scale) and pixel distribution (left vertical scale) relevant to the map in the previous figure. The pixel distribution was evaluated considering a pulse threshold at channel 150.

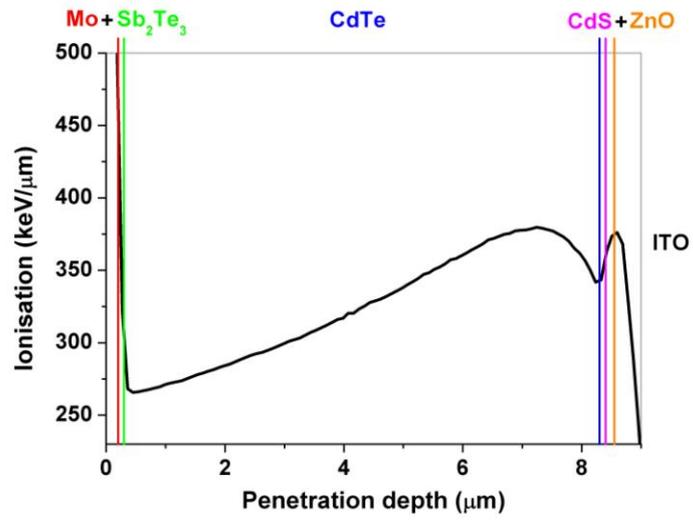

Fig. 8 Simulated energy loss profile in the multilayer cell generated by 3 MeV He ions.

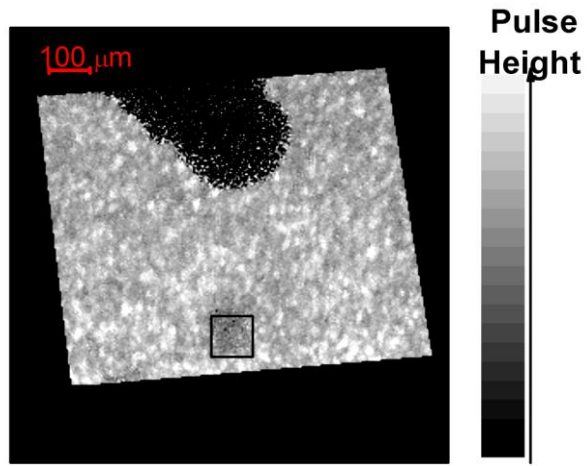

Fig. 9　CCE map of the solar cell after the selective irradiation of the area highlighted in the inset in Fig. 6.

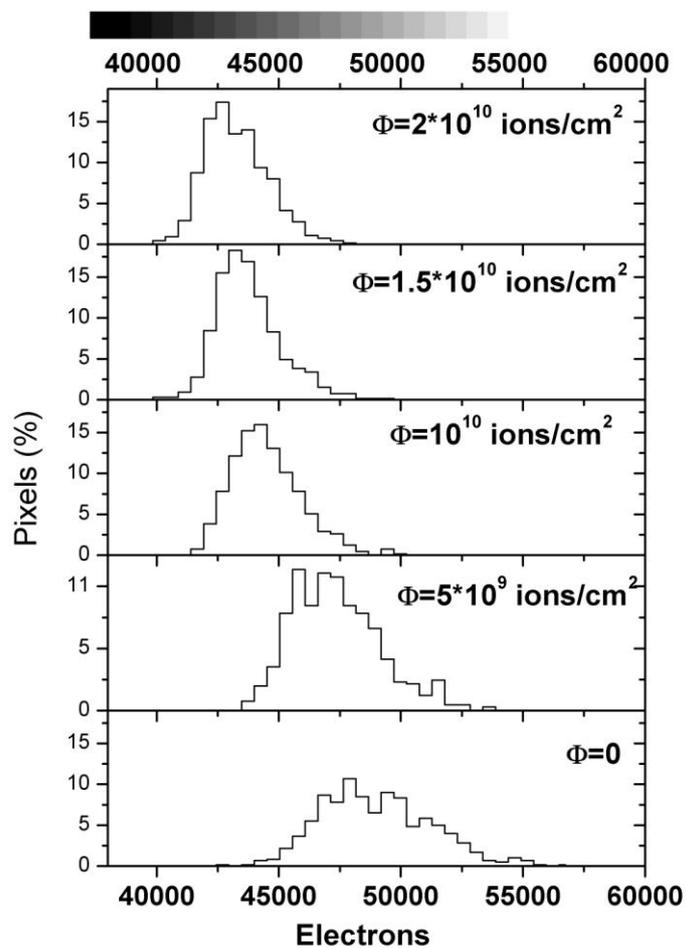

Fig. 10  Spectral evolution of the average IBIC signals recorded after different ion fluences. The measured charge was evaluated by normalising the pulse response by a Si barrier detector used as reference.

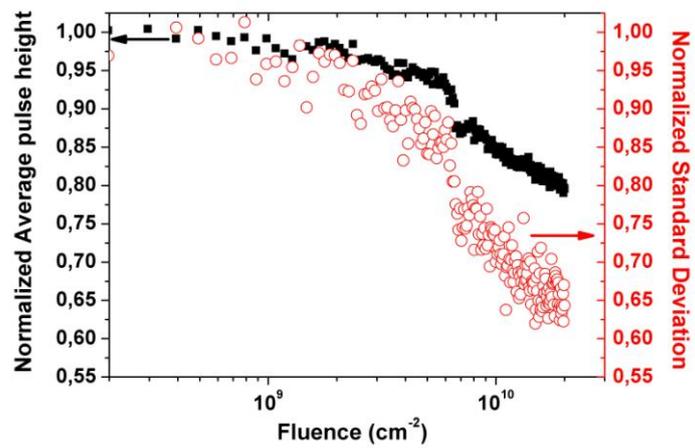

Fig. 11 average CCE (left scale, full square) and standard deviation (right scale, open circle) vs. ion fluence. Data are normalised to the pristine case.

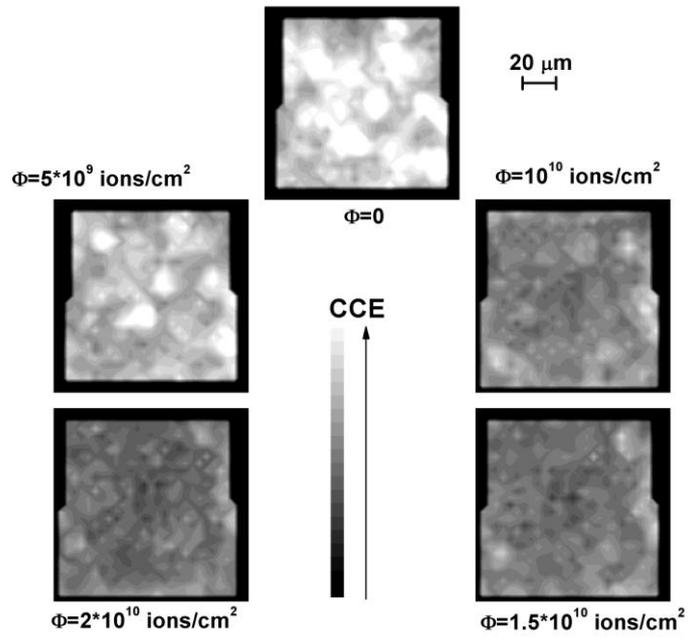

Fig. 12 IBIC maps of the irradiated regions after different fluences.